\documentclass[11pt]{article}
\usepackage[dvips]{graphics}

\def\fw{10cm}
\catcode`\@=11
\ifcase \@ptsize
\or 
\or 
\fi
\advance\textheight by \topskip
\catcode`\@=12

\makeatletter
\@addtoreset{equation}{section}
\makeatother

\gdef\journal#1, #2, #3, 1#4#5#6{{#1~} #2 (1#4#5#6) #3}
\gdef\ibid#1, #2, 1#3#4#5{{#1} (1#3#4#5) #2}
\def\i{{\rm i}}
\def\e{{\rm e}}
\newcommand{\sst}{\scriptstyle}
\catcode`\@=11
\newcommand{\ccite}[1]
{\@ifundefined{b@#1}{\bf ?}{\@nameuse{b@#1}}}
\catcode`\@=12

\newcounter{remark}

\newcommand{\be}{\begin{equation}}
\newcommand{\ee}{\end{equation}}
\newcommand{\bea}{\begin{eqnarray}}
\newcommand{\eea}{\end{eqnarray}}
\newcommand{\bt}{B}
\newcommand{\cP}{\mathcal{P}}

\begin{document}
\begin{center}
{\Large\bf Virial Coefficients of Multispecies Anyons}\\[0.5cm]

{\large
Stefan Mashkevich$^{\mbox{\scriptsize a,b,}}$\footnote{mash@mashke.org},
Jan Myrheim$^{\mbox{\scriptsize b}}$,
K{\aa}re Olaussen$^{\mbox{\scriptsize b}}$}\\[0.3cm]
{\large\it
$^{\mbox{\scriptsize a}}$
Physics Department, Taras Shevchenko Kiev National University,\\
03022 Kiev, Ukraine\\[0.1cm]
$^{\mbox{\scriptsize b}}$
Department of Physics,
The Norwegian University of Science and Technology,\\
N--7034 Trondheim, Norway}
\end{center}

\begin{abstract}
A path integral formalism for multispecies anyons
is introduced, whereby partition functions
are expressed in terms of generating functions of
winding number probability distributions.
In a certain approximation, the equation of state
for exclusion statistics follows.
By Monte Carlo simulation, third-order
cluster and virial coefficients
are found numerically.
\end{abstract}

\section{Introduction}
In two dimensions, a concept of 
statistics of distinguishable particles 
(multispecies statistics, mutual statistics)
arises \cite{W92}, by close analogy with
the concept of fractional (anyonic) statistics \cite{any}.
Multispecies anyons are particles whose wave function picks up a
statistical phase when a particle of one species encircles one of
another species.
Multispecies statistics is interesting per se,
as a novel basic notion in quantum
mechanics, as well as in view of its application to the
quantum Hall effect,
because quasiparticles and quasiholes in quantum Hall systems
do in fact obey such statistics \cite{HJ98}.

A particular model of multispecies anyons in
a high magnetic field has been solved exactly \cite{SSS,SS}. 
The generic case of
free multispecies anyons, however, has not been explored beyond
the (trivial) 2-body system. This is what makes the subject of the
present paper, the focus being on the statistical mechanics of
the free anyon gas.

For the system of more than two identical anyons,
there appears to be no exact
quantum-mechanical solution (multiparticle energy levels
cannot be expressed in terms of single-particle ones
like they can be for bosons and fermions), and even the high-temperature
cluster and virial expansions from
the third order onwards are only known approximately. 
One cannot expect to do any better with multispecies anyons.
Two distinct approaches have been successfully employed
in the treatment of identical anyons: direct numerical computation
of the spectrum \cite{numerics,SJK96} and Monte Carlo
calculation of partition functions \cite{MO93,ASJK98}.
It is the latter that we resort to here.

After introducing the basic definitions and relations in Sec.~2,
we develop, in Sec.~3, a representation of partition functions
in terms of generating functions, which are
Fourier transforms of probability distributions of
winding numbers. This is a direct generalization of the technique
developed in Refs.~\cite{MO93}--\cite{M93}. Then, in Sec.~4,
we introduce an approximation \cite{ASJK98} in which only
two-particle correlations are accounted for.
It yields simple polynomial expressions for the generating functions,
resulting in a virial expansion which
depends on the statistics parameters
at the second order only.
This corresponds to the equation of state for
multispecies exclusion statistics \cite{SS},\cite{mex}.
Section 5 contains the results of Monte Carlo
simulations of the generating functions and the
corresponding mixed virial coefficients.

\section{Definitions and basic relations}

Multispecies anyons are particles in two dimensions, 
characterized by a statistics matrix
$||\alpha_{ab}||$, such that interchanging two particles of species
$a$ supplies the wave function
with a phase factor of $\exp(\i\pi\alpha_{aa})$
and pulling a particle $a$ around a particle $b$
supplies $\exp(2\i\pi\alpha_{ab})$. Due to translation invariance,
the matrix is symmetric. Due to periodicity,
it is enough to consider
$0\le\alpha_{aa}<2$ and $0\le\alpha_{ab}<1$.

The equation of state of a multispecies gas is obtained
as a straightforward generalization of the single-species case.
The fugacity expansion of the grand canonical partition function is
\be
\Xi = \sum_{N_1\ldots N_s} Z_{N_1\ldots N_s} z_1^{N_1}\cdots
z_s^{N_s}\;,
\ee
where $Z_{N_1\ldots N_s}$ is the canonical partition function
of the system containing $N_a$ particles of species $a$ for
each $a=1,\ldots,s$,
and $z_a=\exp(\beta\mu_a)$ are the fugacities
(no chemical equilibrium between species assumed).
The cluster expansion is ($P$ is the pressure, $A$ the area)
\be
\beta P = \frac{\ln \Xi}{A} =
\sum_{N_1\ldots N_s} b_{N_1\ldots N_s} z_1^{N_1}\cdots z_s^{N_s}\;
\ee
and the virial expansion is
\be
\beta P = \sum_{k_1\ldots k_s} A_{k_1\ldots k_s}
\rho_1^{k_1} \cdots \rho_s^{k_s}\;,
\ee
where $\rho_a=z_a\partial (\beta P)/\partial z_a$ are the partial densities.
Relations between partition
functions, cluster and virial coefficients are easily deduced.
The cluster coefficients up to the third order are
\bea
&& Ab_2 = -\frac{1}{2}Z_1^2 + Z_2\;, \qquad
Ab_{11} = -Z_{01}Z_{10} + Z_{11}\;; \nonumber \\
&& Ab_3 = \frac{1}{3}Z_1^3 - Z_1Z_2 + Z_3\;, \label{b3} \qquad
Ab_{21} = Z_{01}Z_{10}^2 - Z_{10}Z_{11} - Z_{01}Z_{20} + Z_{21}\;, \\
&& Ab_{111} = 2 Z_{001} Z_{010} Z_{100} - Z_{100} Z_{011} -
Z_{010} Z_{101} - Z_{001} Z_{110} + Z_{111}\;, \nonumber
\eea
and the virial coefficients are
\bea
&& A_2 = -\bt_2\;, \qquad
A_{11} = -\bt_{11}\;; \nonumber \\
&& A_3 = 4\bt_2^2 - 2\bt_3\;, \label{A111-b} \qquad
A_{21} = \bt_{11}^2 + 4\bt_{11}\bt_{20} - 2\bt_{21}
\;,\label{A21-b} \\
&& A_{111} = 2 \bt_{011} \bt_{101} +
 2 \bt_{011} \bt_{110} +
 2 \bt_{101} \bt_{110} - 2 \bt_{111}\;, \nonumber 
\eea
where
\bea
\bt_{N_1\ldots N_s} & = & \frac{b_{N_1\ldots N_s}}
{b_{100\ldots0}^{N_1} b_{010\ldots0}^{N_2} \cdots b_{0\ldots01}^{N_s}}\;.
\eea
Dimensionless virial coefficients are
\be
a_{k_1\ldots k_s} =
\frac{A_{k_1\ldots k_s}}{\lambda^{2(k_1+\cdots+k_s-1)}}\;,
\ee
$\lambda=\sqrt{2\pi\beta/m}$ being the thermal wavelength.
(Cf.~\cite{SSS,SS}; 
the cluster coefficients here contain an extra factor of $1/A$
as compared to the notation used there.)

\section{The generating functions and their connected parts}

We now wish to represent the partition function $Z_{N_1\ldots N_s}$
as a path integral over closed paths
in configuration space. Since
particles of the same species are indistinguishable, a path may induce
a permutation of such particles. This breaks up all paths into
classes which constitute a representation
of the ``colored permutation group'',
a direct product of permutation groups
$S_{N_1}\otimes\ldots\otimes S_{N_s}$.
To a conjugation class $\cP$ of elements
of this group there corresponds a set
$\{ \nu_L^a \::\: a=1,\ldots,s, \:L=1,\ldots,\infty \}$
such that $\{ \nu_L^a \}$ for a fixed $a$ label a partition
of $N_a$, in the sense that $\sum_{L=1}^\infty \nu_L^a L = N_a$.
That is, $\nu_L^a$ is the number of times $L$ figures in the partition
of $N_a$; in other words, the permutation of $N_a$ elements is represented
by cycles, and $\nu_L^a$ is the number of cycles of length $L$.
The expression for the partition function then is
\be
Z_{N_1\ldots N_s} = \sum_\cP F_\cP \prod_{a=1}^s
\prod_{L=1}^\infty \frac{1}{\nu_L^a!}
\left( \frac{Z_1(L\beta)}{L} \right)^{\nu_L^a} \;,
\label{Z}
\ee
where the sum is over all conjugation classes $\cP$ of the group,
the double product is the contribution from
$\cP$ to the partition function of $(N_1+\cdots+N_s)$ 
multispecies bosons, and $F_\cP$ comes
from the interaction. By a reasoning completely analogous to
that in Refs.~\cite{MO93,M93}, the central idea being
that the effect of the anyonic ``statistics interaction''
is to supply a path with an extra phase coming from 
the mutual winding of particles, one concludes that 
$F_\cP$ is the generating function of
winding number probabilities, given by
\be
F_\cP = \sum_{\{Q_{jk}\}}
P_\cP (\{Q_{jk}\})
\exp\big(\!-\!\i\pi\sum_{j<k}\alpha_{a_ja_k}Q_{jk}\big)\;,
\label{FP}
\ee
where $j$ or $k$ numbers all the particles, from 1 to $\sum_a N_a$;
$a_j$ is the species that particle $j$ belongs to;
and $P_\cP (\{ Q_{jk} \})$ is the probability
that for a path inducing a permutation from class $\cP$,
the pairwise winding numbers form the set $\{Q_{jk}\}$,
the distribution of paths being the same as for free bosons
at inverse temperature $\beta$.

The winding number $Q_{jk}$ is defined here as $\phi_{jk}/\pi$ where
$\phi_{jk}$ is the angle by which the radius vector
${\bf r}_j-{\bf r}_k$ rotates.
It is an even integer if particles $j$ and $k$
regain their initial positions, an odd integer if they get exchanged,
but generally a fractional number if one or both particles
exchange positions with others of the same species. An individual
$Q_{jk}$ is therefore a continuous variable in general, which 
seemingly makes the sum in (\ref{FP}) somewhat ill-defined.
However, it can be observed that

(i) if particles $j_1, \ldots, j_m$ make up a cycle (i.e., they are all
of species $a$ and get cyclically permuted by the path), then it is
only the ``intracycle winding number''
\be
Q_{j_1\ldots j_m} = \sum_{p<q=1}^m Q_{j_p j_q}
\label{Qintra}
\ee
that figures (multiplied by $\alpha_{aa}$) in the phase factor
in Eq.~(\ref{FP});

(ii) if particles $j_1, \ldots, j_m$ (species $a$) make up one cycle and
$k_1, \ldots, k_n$ (species $b$) make up another cycle, then it is
only the ``intercycle winding number''
\be
Q_{(j_1\ldots j_m)(k_1\ldots k_n)} = \sum_{p=1}^m \sum_{q=1}^n Q_{j_p k_q}
\label{Qinter}
\ee
that figures (multiplied by $\alpha_{ab}$) in the phase factor.

Now, $Q_{j_1\ldots j_m}$ is always an integer (even/odd for $m$ odd/even)
and $Q_{(j_1\ldots j_m)(k_1\ldots k_n)}$ is always an even integer.
We can then sum over these numbers, with the
corresponding probabilities, rather than over individual $Q_{jk}$'s.
Classes of paths characterized by their
winding numbers form representations
of a so-called colored braid group (paths within the same class
can be continuously deformed one into another).

We will, as mentioned above, characterize permutations
by their corresponding sets of cycles, and explicitly
write out $F_\cP$ in terms of cycles, as
\arraycolsep 0.0em
\renewcommand{\arraystretch}{0.6}
$F{}
\begin{array}{ccccccc}
\sst 1 & \sst \ldots & \sst 1 & \sst \ldots & \sst s & \sst \ldots & \sst s \\
\sst L_{11} & \sst \ldots & \,\sst L_{1\nu^1} & \sst \ldots & \,\sst L_{s1}
& \sst \ldots & \,\sst L_{s\nu^s}
\end{array}
$,
\arraycolsep 0.15em
\renewcommand{\arraystretch}{1}
where
$\{ L_{a1},\ldots,L_{a\nu^a} \}$ is the set of lengths of cycles pertaining
to species $a$. 
(The species sequence, shown here for simplicity as $1,\ldots,s$,
may in fact miss some members and may not come in order.)
E.g., $F{}^{aab}_{213}$ is the generating function
for paths involving three particles of species $a$ and three particles
of species $b$, inducing a permutation which interchanges two of
the species $a$ particles (cycle of length 2) and cyclically permutes
all three particles of species $b$.

The cluster coefficients are expressed in terms of the $F$-functions
via Eqs.~(\ref{b3}) and (\ref{Z}). 
In the thermodynamic limit $A\to\infty$, which we will be interested in,
the single-particle partition function is
$Z_1(\beta)=A/\lambda^2 \propto \beta^{-1}$,
so that $Z_1(L\beta)=Z_1(\beta)/L$. Denoting, for brevity,
$Z_1 \equiv Z_1(\beta)$, one gets the following relations:
\bea
&& \lambda^2 b_{2} = \frac12 G{}_{11}^{11} + 
        \frac14 G{}_{2}^{1} \;, \qquad
\lambda^2 b_{11} = G{}_{11}^{12} \;;\label{b2ord} \\
&& \lambda^2 b_{3} = \frac16 G{}_{111}^{111} + 
        \frac14 G{}_{21}^{11} + \frac19 G{}_{3}^{1} \;, \qquad
\lambda^2 b_{21} =
\frac12 G{}_{111}^{112} +
	\frac14 G{}_{21}^{12}\;, \qquad
\lambda^2 b_{111} =
G_{111}^{123}\;,\label{b3ord}
\eea
where
\bea
&& G{}_{2}^{a} = F{}_{2}^{a}\;, \qquad
G{}_{11}^{ab} = [F{}_{11}^{ab} - 1]Z_1\;;\label{G2ord} \\
&& G{}_{3}^{a} = F{}_{3}^{a}\;, \qquad
G{}_{21}^{ab} = [F{}_{21}^{ab} - F{}_{2}^{a}]Z_1\;, \qquad
G{}_{111}^{abc} = [F{}_{111}^{abc} -
F{}_{11}^{ab} - F{}_{11}^{ac} - F{}_{11}^{bc} + 2]Z_1^2\;\label{G3ord}
\eea
are the ``connected parts'' of the $F$'s, in a sense to be
clarified below.

{}From the single-species case, it is known \cite{ASJK98,DHOdV} that
\be
G{}_{L}^{a} = F{}_{L}^{a} = \prod_{k=1}^{L-1} \left(1-\frac{L}{k}\alpha_{aa}\right)\;,
\ee
so, in particular,
\be
G{}_{2}^{a} = 1-2\alpha_{aa}\;, \qquad
G{}_{3}^{a} = 1-\frac92\alpha_{aa}+\frac92\alpha_{aa}^2 \;.
\ee
Also, from the exact two-anyon partition function it can be derived
that $G{}_{11}^{aa} = \alpha_{aa}(\alpha_{aa}-1)$. Since this comes from
paths of a single pair of particles without permutation, the same result
holds for distinguishable particles as well:
\be
G{}_{11}^{ab} = \alpha_{ab}(\alpha_{ab}-1)\;.
\ee
Hence, in the second order both the single-species \cite{ASWZ85}
and mixed cluster and virial coefficients are known exactly:
\bea
\lambda^2 b_{2}(\alpha_{11}) & = & 
- a_{2}(\alpha_{11})
= \frac12 (1-\alpha_{11})^2 - \frac14\;,\label{ba2}\\
\lambda^2 b_{11}(\alpha_{12}) & = & 
- a_{11}(\alpha_{12})
= \alpha_{12}(\alpha_{12}-1)\;,\label{ba11}
\eea
Note that $b_{11}(\alpha_{12}) = b_{2}(\alpha_{12}) + b_{2}(1+\alpha_{12})$,
reflecting the fact that the basis wave functions of two distinguishable
anyons can be chosen to consist of symmetric and antisymmetric functions
multiplied by the same $\alpha_{12}$ dependent anyonic prefactor as for
identical anyons. 

\section{The two-particle approximation}

The physical meaning of the $G$-functions becomes apparent as soon
as they are cast explicitly in terms of winding number probabilities.
Consider, e.g., $G_{21}$, which depends on two statistics
parameters, $\alpha_{11}$ and $\alpha_{12}$.
(Once we start explicitly writing down the species-dependent
statistics parameters, we drop the species-related superscripts
on the $G$'s as long as this causes no ambiguity.)
According to the definitions (\ref{G3ord}) and (\ref{FP})--(\ref{Qinter}),
\be
G_{21}(\alpha_{11},\alpha_{12})=
\sum_{Q_{12}Q_{(12)(3)}} P_{Q_{12}Q_{(12)(3)}}
\left( \e^{\i\pi [\alpha_{11}Q_{12} + \alpha_{12}Q_{(12)(3)}]}
- \e^{\i\pi \alpha_{11}Q_{12}}\right)\;.
\ee
Terms with $Q_{(12)(3)}=0$ do not contribute to the sum.
In other words, only those paths do contribute
where the 2-cycle formed by the first two particles,
of species 1, actually winds around the 1-cycle
which is the third particle, of species 2.
Generally, a path which is disconnected in the sense that 
there exist two subsets of cycles such that the path
involves no winding of any cycle in the first subset
around any cycle in the second one, will not contribute
to its corresponding $G$-function.
It is in this sense that we call the latter the
connected part of the $F$-function.

The probability for a path involving two distinguishable particles
to have a nonzero winding number is inversely proportional
to $Z_1$---i.e., it tends to zero in the thermodynamic
limit.
Indeed, a single-particle path, on average, covers
an area proportional to $\lambda^2$, and $\lambda^2/A = Z_1^{-1}$
is the probability of two such paths coming close to each other.
This makes it sensible to suggest an approximation
whose essence is neglecting more than two-particle
correlations. Specifically, this is an approximation
of uncorrelated minimal pairwise windings, in which:
(i) any intercycle winding is replaced with
all possible combinations of pairwise windings 
between one particle belonging to one of the cycles
and another one belonging to the other (i.e., the
fractional pairwise winding numbers are replaced with
integers); (ii) only minimally connected paths are taken
into account (e.g., for three particles, if $Q_{12} \ne 0$
and $Q_{13} \ne 0$, which is enough for connectedness,
$Q_{23}$ is assumed to be equal to zero);
(iii) windings of different
pairs are supposed to be mutually uncorrelated
[which is, of course, an exact statement with respect
to pairs (12) and (34), but not with respect to (12) and (13)].

Since the generating function for combined probabilities
of uncorrelated events factorizes, any $G$-function
in this approximation (to be denoted by $\tilde G$)
is a sum of products of $G_L$'s
and $G_{11}$'s with different arguments, i.e.,
a polynomial in the $\alpha_{ab}$'s. In particular,
\be
\tilde G_{21}(\alpha_{11}, \alpha_{12})
= 2\,G_2(\alpha_{11})G_{11}(\alpha_{12})
= 2(1-2\alpha_{11})\alpha_{12}(\alpha_{12}-1)\;,
\label{G21pol}
\ee
representing a distribution of the
2-cycle windings, $G_2$, and an uncorrelated distribution 
of windings of the third particle around either the
first or the second one, $2G_{11}$.

Contributing to the approximation for $G_{111}$
are paths with two, but not three of the winding numbers
nonvanishing; therefore,
\bea
\tilde G_{111}(\alpha_{12},\alpha_{13},\alpha_{23}) & = &
G_{11}(\alpha_{12})G_{11}(\alpha_{13}) +
G_{11}(\alpha_{12})G_{11}(\alpha_{23}) +
G_{11}(\alpha_{13})G_{11}(\alpha_{23}) \nonumber \\
& = & 
\alpha_{12}(\alpha_{12}-1)\alpha_{13}(\alpha_{13}-1)+
\alpha_{12}(\alpha_{12}-1)\alpha_{23}(\alpha_{23}-1) \nonumber \\
&& {} +
\alpha_{13}(\alpha_{13}-1)\alpha_{23}(\alpha_{23}-1)\;.
\eea
One might be tempted to surmise that this expression is in fact
exact in the thermodynamic limit, the contribution of
paths with the third winding number nonvanishing
containing another factor of $Z_1^{-1}$.
That is of course not the case.
If both $Q_{12}$ and $Q_{13}$ are nonzero,
particles 2 and 3 are already close to each other, since
they are both close to particle 1.
Still, it is a sensible starting point, as indeed confirmed
by the simulations. (For a single species,
a straightforward diagrammatic
interpretation has been adduced \cite{ASJK98}.
Its generalization to many species is possible, as is a certain
improvement of the approximation itself \cite{SJKf}.)

Remarkably, in this polynomial approximation,
starting with the third order, 
all the virial coefficients turn out to be
statistics independent (in particular, all the mixed
virial coefficients vanish).
Introducing the $H$-functions
which are the deviations from the approximation in question,
\be
G_\cP = \tilde G_\cP + H_\cP\;,
\ee
and using the explicit expressions for the $\tilde G$'s,
one gets, in the third order,
\bea
a_3(\alpha_{11}) &=& \frac{1}{36} -
\frac{1}{3}H_{111}(\alpha_{11},\alpha_{11},\alpha_{11}) \;, \label{A3} \\
a_{21}(\alpha_{11}, \alpha_{12}) &=& 
- \frac12 H_{21}(\alpha_{11},\alpha_{12})
- H_{111}(\alpha_{11},\alpha_{12},\alpha_{12}) \;, \label{A21} \\
a_{111}(\alpha_{12},\alpha_{13},\alpha_{23}) &=& 
-2 H_{111}(\alpha_{12},\alpha_{13},\alpha_{23}) \; \label{A111}
\eea
(in the first formula, it has been taken into account
that $H_{21}(\alpha_{11},\alpha_{11}) \equiv 0$, since
for a single species, the polynomial approximation
for $G_{21}$ is known \cite{ASJK98,M93} to be exact),
and likewise in the higher orders.
This is the simplest possible way in which the equation of state
can interpolate between the Bose and Fermi limits.
Absence of corrections of the third and higher orders
in density is of course a consequence of the multiparticle
effects being neglected.
The corresponding equation of state
is characteristic of multispecies exclusion statistics \cite{SS}.
Thus, as well as for a single species \cite{ASJK98},
the thermodynamics of exclusion statistics is reproduced
by anyons as long as only two-particle
correlations are taken into account.

\section{Monte Carlo simulation and results}

\begin{figure}
\begin{center}
\resizebox{\fw}{!}{
\includegraphics{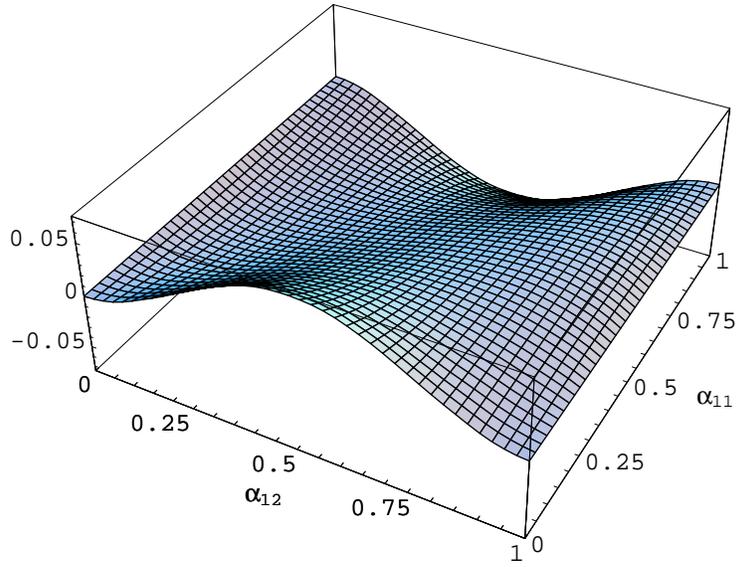}
}

\caption
{
$H_{21}(\alpha_{11},\alpha_{12})$.
}

\label{fig1}
\end{center}
\end{figure}

\begin{figure}
\begin{center}

\begin{tabular}{cc}
\scalebox{.6}{
\includegraphics{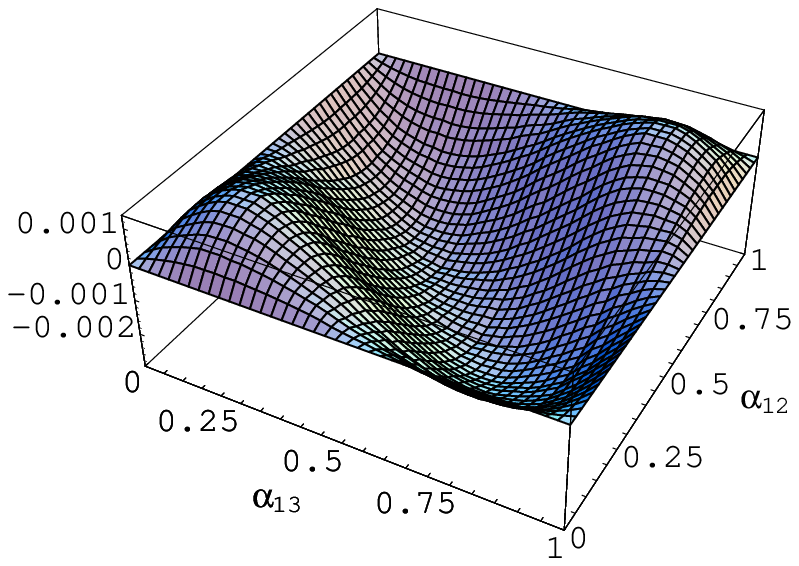}
}
&
\scalebox{.6}{
\includegraphics{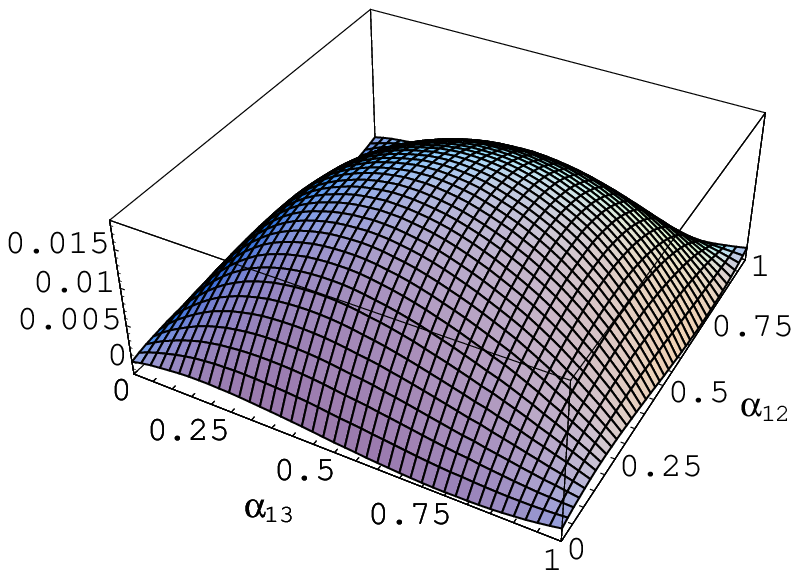}
}
\\
(a) & (b) \\
\multicolumn{2}{c}{
\scalebox{.6}{
\includegraphics{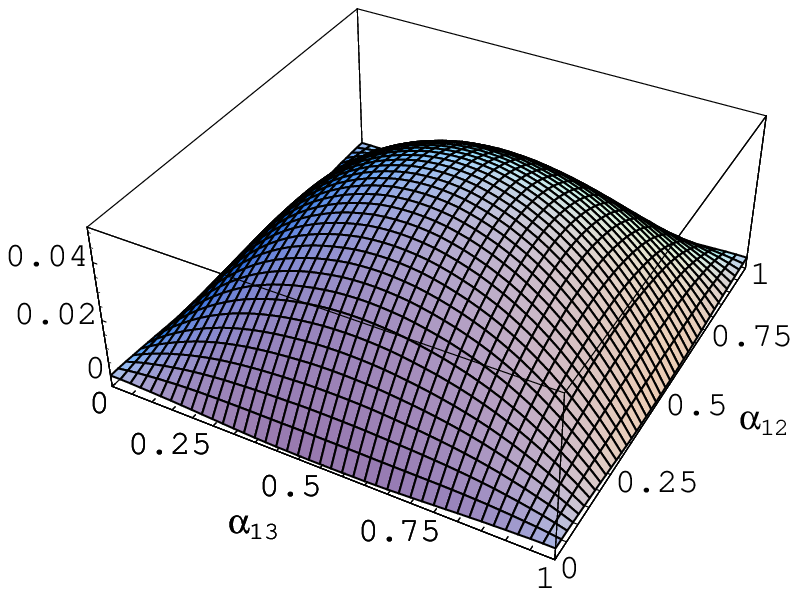}
}
} \\
\multicolumn{2}{c}{(c)}
\end{tabular}

\caption
{
The mixed third virial coefficient
$a_{111}(\alpha_{12},\alpha_{13},\alpha_{23})
= -2H_{111}(\alpha_{12},\alpha_{13},\alpha_{23})$
at (a) $\alpha_{23}=0$, (b) $\alpha_{23}=0.1$, (c) $\alpha_{23}=0.5$.
}

\label{fig2}
\end{center}
\end{figure}

We now turn to a direct numerical computation of
the generating functions, and thence of the 
cluster and virial coefficients.
To that end, we Monte Carlo simulate random paths
inducing a given permutation,
with the thermal distribution valid for free bosons.
For each path, we calculate the intracycle and
intercycle winding numbers as defined by
Eqs.~(\ref{Qintra}), (\ref{Qinter}).
The resulting winding number probabilities
are substituted into the definitions of the generating functions.

\begin{figure}
\begin{center}
\resizebox{\fw}{!}{
\includegraphics{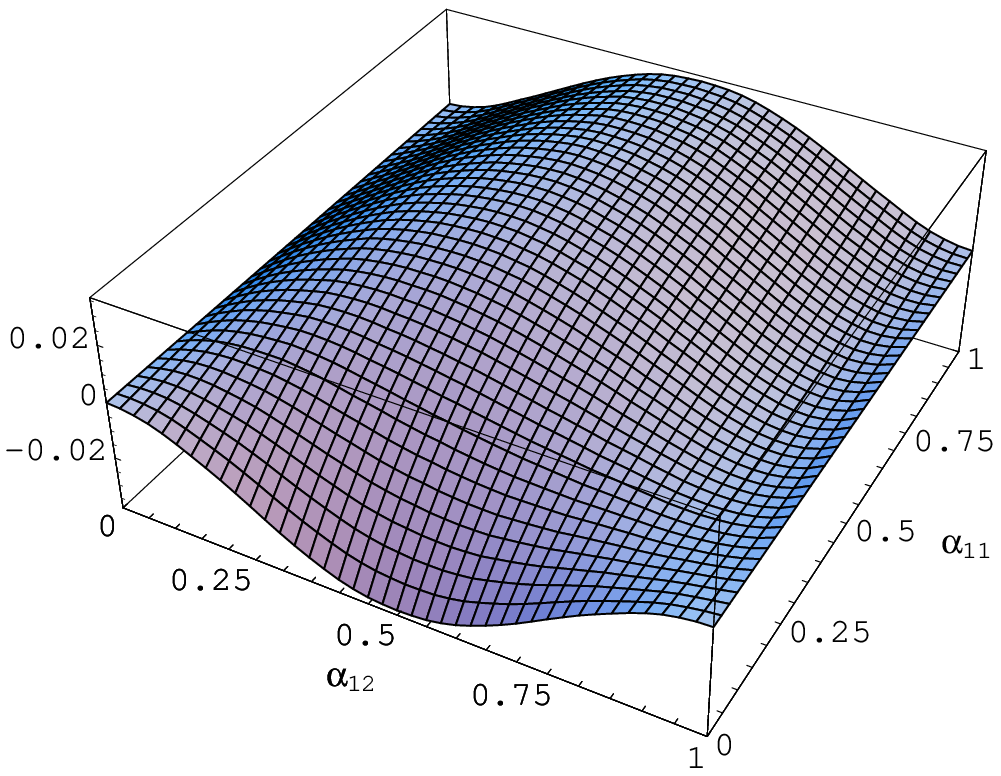}
}

\caption
{
The mixed third virial coefficient
$a_{21}(\alpha_{11},\alpha_{12})$.
}

\label{fig3}
\end{center}
\end{figure}

We are interested in the thermodynamic limit, in which
$Z_1^{-1}\to0$; but in practice,
the calculations have to be performed at a finite value
of $Z_1^{-1}$, determined by the choice of $\beta$.
(Both $m$ and $A$ can be scaled to unity, then $Z_1^{-1}=2\pi\beta$.)
A compromise is necessarily involved: This value
cannot be too big lest finite-size effects emerge,
but it cannot be too small because, as explained above,
the number of useful events tends to zero together with it.
We have chosen such values (different for different partitions)
that typically several paths out of a thousand contribute
to the connected parts.

For the (21) partition, $32.1\times10^6$ paths were generated
at $\beta=0.01$ ($Z_1^{-1}=0.0628$).
Figure~\ref{fig1} shows the correction $H_{21}(\alpha_{11},\alpha_{12})$
to the polynomial approximation $\tilde G_{21}$, Eq.~(\ref{G21pol}).
It vanishes, as it should, for $\alpha_{12}=0$ and $\alpha_{12}=1$
(no statistical interaction between the two species), as well
as for $\alpha_{11}=\alpha_{12}$ (see above). Also, 
like $\tilde G_{21}$ itself, it is antisymmetric with
respect to $(\alpha_{11},\alpha_{12}) \mapsto (1-\alpha_{11},1-\alpha_{12})$,
since the total winding number is always odd.
The maximum is $H_{21}(0,0.5) \simeq 0.071$, which is to be compared
with the maximum of $\tilde G_{21}(0,0.5) = 0.500$; thus, the relative
error of the two-particle approximation is about 15\%.
The absolute error of the MC data themselves, which can be
estimated by looking at the imaginary part of the simulated
function, never surpasses $0.001$.

For the (111) partition, $340\times10^6$ paths were generated
at $\beta=0.03$ ($Z_1^{-1}=0.188$). 
We illustrate our results with plots of
$a_{111}(\alpha_{12},\alpha_{13},\alpha_{23})
= -2H_{111}(\alpha_{12},\alpha_{13},\alpha_{23})$
at fixed values of $\alpha_{13}=0$, $0.1$, and $0.5$,
on Fig.~\ref{fig2} (a), (b), (c), resp.
The first of the three cases, where winding numbers of
one of the pairs do not contribute, reflects
the three-particle correlation in the windings of
the pairs (12) and (13). The extreme value,
$H_{111}(0.5,0.5,0) \simeq 0.00138$,
is to be compared with the conditional maximum
$\tilde G_{111}(0.5,0.5,0) = 0.06250$.
The approximation of uncorrelated windings per se
is, therefore, accurate up to 2\%.
When all three parameters are nonzero, it is
not only the above-mentioned correlation
but also any paths with all three
winding numbers nonvanishing
that start to contribute to the correction,
making the latter bigger by almost an order of magnitude.
The extreme value is $H_{111}(0.5,0.5,0.5) \simeq -0.0252$
vs.~$\tilde G_{111}(0.5,0.5,0.5) = 0.1875$;
the error is $0.00025$ at most.
Correspondingly,
the extreme value of the three-species mixed virial coefficient
is $a_{111}(0.5,0.5,0.5) \simeq 0.0504$.

Our next result, Fig.~\ref{fig3}, is the two-species
mixed third virial coefficient $a_{21}(\alpha_{11},\alpha_{12})$,
determined from Eq.~(\ref{A21}).
It includes an antisymmetric [with respect to
$(\alpha_{11},\alpha_{12}) \mapsto (1-\alpha_{11},1-\alpha_{12})$]
part, $H_{21}$, and a symmetric one, $H_{111}$.
The latter being mostly negative, the virial
coefficient is predominantly positive;
however, its extreme negative value,
$a_{21}(0, 0.5) \simeq -0.0368$ is slightly
bigger by magnitude than its extreme positive value,
$a_{21}(1, 0.5) \simeq 0.0343$.

As a final check, we have computed the single-species
third-order virial coefficient $a_3(\alpha_{11})$,
and found the result to be consistent with the
known formula \cite{SJK96},
\be
a_3(\alpha) = \frac{1}{36} + \frac{1}{12\pi^2}\sin^2 \pi\alpha
- (1.65\pm0.01)\times 10^{-5} \sin^4 \pi\alpha \;,
\ee
within the error margin of $0.0002$.
(The magnitude of the error is bigger than that
of the $\sin^4$ term, which cannot therefore be seen here.)

The next logical step is to try to infer
the approximate analytic formulas for
the mixed virial coefficients,
either as fits to the MC data or by somehow taking
into account multiparticle correlations in the winding
number distributions.
It is possible to make some progress on both avenues;
results will be reported elsewhere \cite{SJKf}.

\end{document}